\documentclass[letterpaper, 10 pt, conference]{ieeeconf}
\IEEEoverridecommandlockouts
\usepackage{cite}
\usepackage{amsmath}
\usepackage{amssymb}
\usepackage{amsfonts}
\usepackage{algorithmic}
\usepackage{graphicx}
\usepackage{textcomp}
\usepackage[table]{xcolor}
\usepackage{balance}
\usepackage{subcaption}
\usepackage{bm}
\usepackage{booktabs}

\usepackage{algorithm2e}
\RestyleAlgo{ruled}
\usepackage{algorithmic}

\makeatletter
\newcommand{\nosemic}{\renewcommand{\@endalgocfline}{\relax}}
\newcommand{\dosemic}{\renewcommand{\@endalgocfline}{\algocf@endline}}
\newcommand{\pushline}{\Indp}
\newcommand{\popline}{\Indm\dosemic}
\let\oldnl\nl
\newcommand{\nonl}{\renewcommand{\nl}{\let\nl\oldnl}}
\makeatother

\title{\LARGE \bf
Hospital-Agnostic Image Representation Learning in Digital Pathology
}

\author{Milad Sikaroudi$^{1}$, Shahryar Rahnamayan$^{1,2}$, H.R. Tizhoosh$^{1,3,*}$
\thanks{$^{1}$M. Sikaroudi, S. Rahnamayan, H.R. Tizhoosh are  members of the  Laboratory  for  Knowledge  Inference  in  Medical  Image  Analysis (Kimia Lab), University of Waterloo, Waterloo, ON, Canada, 
{(e-mail: \{msikaroudi, s5rahnamayan, tizhoosh\}@uwaterloo.ca}).}%
\thanks{$^{2}$ S. Rahnamayan, Electrical, Computer and Software Engineering, Ontario Tech University, Oshawa, ON, Canada}
\thanks{$^{3}$ H. R. Tizhoosh, Artificial Intelligence and Informatics, Mayo Clinic, Rochester, MN, USA {(e-mail: tizhoosh.hamid@mayo.edu})}%
\thanks{$^{*}$Corresponding author: H.R. Tizhoosh.}
}
\usepackage{eso-pic}

\begin{document}

\AddToShipoutPictureBG*{%
  \AtPageUpperLeft{%
    \setlength\unitlength{1in}%
    \hspace*{\dimexpr0.5\paperwidth\relax}
    \makebox(0,-0.75)[c]{\small Accepted for presentation at the 44th Annual International Conference of the IEEE Engineering in Medicine and Biology Society (EMBC'22)}
    }}

\bstctlcite{IEEEexample:BSTcontrol}

\maketitle

\begin{abstract}

Whole Slide Images (WSIs) in digital pathology are used to diagnose cancer subtypes. The difference in procedures to acquire WSIs at various trial sites gives rise to variability in the histopathology images, thus making consistent diagnosis challenging. These differences may stem from variability in image acquisition through multi-vendor scanners, variable acquisition parameters, and differences in staining procedure; as well, patient demographics may bias the glass slide batches before image acquisition. These variabilities are assumed to cause a domain shift in the images of different hospitals. It is crucial to overcome this domain shift because an ideal machine-learning model must be able to work on the diverse sources of images, independent of the acquisition center. A domain generalization technique is leveraged in this study to improve the generalization capability of a Deep Neural Network (DNN), to an unseen histopathology image set (i.e., from an unseen hospital/trial site) in the presence of domain shift. According to experimental results, the conventional supervised-learning regime generalizes poorly to data collected from different hospitals. However, the proposed hospital-agnostic learning can improve the generalization considering the low-dimensional latent space representation visualization, and classification accuracy results.
\end{abstract}

\section{INTRODUCTION}

Hematoxylin \& Eosin (H\&E) stained slides of biopsy samples on glass slides are widely used by pathologists to examine human tissue under the microscope \cite{chen2014histological}. Through the digitization of glass slide into Whole Slide Images (WSIs) using advanced scanners, the door to computer-aided image analysis and machine learning has been opened for the pathology field. Due to the emergence of deep learning and thanks to the existence of large datasets, image classification has experienced significant breakthroughs in closing the so-called semantic gap. Additionally, Convolutional Neural Networks (CNNs) have been successful in capturing complex tissue patterns, and are widely used in biomedical imaging for cancer diagnosis \cite{jimenez2017analysis}. However, because of the limited availability of expert-labeled training data \cite{tizhoosh2018artificial}, and since these models are still not capable of generalizing beyond training data \cite{ruder2017transferlearning}, their clinical utility is rather questionable for computational pathology. 

\emph{Transfer learning} is utilized to overcome the mentioned limitation. It deals with re-purposing previously learned abstract information in a new context, similar to humans in the way that they do not learn everything from scratch and instead use and transfer their knowledge from previously learned areas to new domains and tasks. In histopathology, the recurring histomorphologic patterns are a hallmark of organizing diseases into meaningful subgroups by pathologists, hence transfer learning can be leveraged \cite{sikaroudi2020supervision}. But, the problem is the assumption that the source and target domains are IID (Identically Independent Distribution) while disregarding OOD (Out of Distribution) scenarios that occur frequently in reality, leading to DNNs utter failures \cite{torralba2011unbiased}. The discrepancy between the source and target domain distributions is known as \textit{domain shift} \cite{quionero2009dataset}.
According to Yagi et al. \cite{yagi2011color}, the domain shift in histopathology usually occurs when WSIs are acquired at different trial sites due to differences in slide preparation, staining procedure, and scanner characteristics as well as biases in developing AI models \cite{dehkharghanian2021biased}. Overall, correct diagnosis is a critical task in histopathology which can directly influence the treatment outcome. Therefore, for the sake of patient well-being, it is important to train robust models invariant to the domain shift (a.k.a being ``\emph{hospital-agnostic}''). Increasing the diversity of training data or stain normalization \cite{reinhard2001color} may alleviate the domain shift problem; however, these methods are rather subjective and mostly expensive solutions. Hence, this study will attempt to show how Multi-Domain Learning (MDL) regimes, specifically a Domain Generalization (DG) technique, namely Model Agnostic Learning of Semantic Features (MASF) \cite{dou2019domain} can be leveraged to rectify the problem  \cite{sikaroudi2021magnification}. 

\section{Literature Review}
In MDL, the aim is to train a single model which is effective for multiple known domains \cite{bilen2017universal}. Domain Adaptation (DA) and DG are two major categories of MDL distinguishing in terms of their assumptions. The DG is more realistic in comparison to DA because DG assumes that there is not any access to target domain data at all. A promising research direction in DG is to leverage the learning regime of Model Agnostic Meta Learning (MAML) \cite{finn2017model}. In what follows, MASF will be introduced.

\subsection*{Model Agnostic Learning of Semantic Features}

Inspired by the MAML approach, MASF learns a latent space representation suitable for generalization to an unseen target domain. MASF is a model agnostic learning scheme that can be used for performing different tasks. If the classification is the task at hand, MASF consists of some layers of feature extraction, classification, and metric embedding.

MASF contains three different loss functions for learning a discriminative embedding space. These loss functions are a cross-entropy loss for supervised learning, a Kullback-Leibler  (KL) divergence loss for domain alignment purpose, and a metric loss to promote domain-independent class-specific cohesion and separation of instances.

Consider that a CNN, consisting of some feature extraction and classification layers, is being used as the model in a MASF framework. Then, $G_\psi$ sub-network is the feature extractor that ends up in latent space representation, and $S_\theta$ sub-network is the classification layers of the model of this CNN. A metric learning module, $M_\phi$ also comes after $G_\psi$ which can be a triplet loss or contrastive loss functions. 

At each iteration of the training, a batch of the source domain dataset ($\mathcal{D}$) is split into meta-train and meta-test batches, indicating by $\mathcal{D}_\text{tr}$, and $\mathcal{D}_\text{te}$, respectively. Then, $G_\psi$, $S_\theta$, and $M_\phi$ sub-networks are updated.

\section{METHODOLOGY}

\subsection{Preprocessing}
Consider there are $K$ sets of WSI repositories which have been collected from different trial sites as $\mathcal{H}_{s} = \{H_k\}^K_{k=1}$. Once the tissue regions from the background regions are segmented (using Otsu algorithm \cite{otsu1979threshold}), a dataset of patches is created, i.e.,  $\mathcal{H}_{p} = \{H_k\}^K_{k=1}$, to feed into the upcoming CNN. As it is shown in Algo. \ref{alg:hospital-agnostic}, the leave-one-\emph{hospital}-out technique is used, i.e., training on the repositories of $K-1$ trial sites and testing on one hold-out unseen repository. So, $\mathcal{H}_p$ is split into $\mathcal{H}_{\text{external}} =  \{H_k\}$ and $\mathcal{H}_{\text{internal}} =  \{H_i\}^K_{i\neq k}$. 

\subsection{Hospital-Agnostic Learning Regime}
Consider $K$ training source domain trial sites, denoted by $\mathcal{H}_{\text{internal}} =  \{H_i\}^K_{i\neq k}$. In every iteration, the source domain trial sites are split into meta-train and meta-test, denoted by $\mathcal{H}_\text{tr}$ and $\mathcal{H}_\text{te}$, respectively. Three different loss functions are utilized, as in MASF. We describe these loss functions below.

\subsubsection{Cross-Entropy Loss}
\begin{align}
&\mathcal{L}_\text{ce}(\mathcal{H}_\text{tr}; \psi, \theta) := \\
\nonumber
&~~~~~~~~~~\frac{-1}{|\mathcal{H}_\text{tr}|} \sum_{\mathcal{H} \in \mathcal{H}_\text{tr}} \frac{1}{|\mathcal{H}|} \sum_{(\bf{x},y) \in \mathcal{H}} \sum_{c=1}^C \mathbb{I}(y = c) \log \mathbb{I}(\widehat{y} = c), \label{equ:cross_entropy}
\end{align}
$|\cdot|$ indicates the cardinality of the set, $C$ is the number of classes, $y$ is the true label for $\bf{x}$, $\widehat{y}$ is the predicted label for $\bf{x}$, and $\mathbb{I}(\cdot)$ is the indicator function which is one when its condition is met, and zero otherwise. \newline

\subsubsection{Hospital-Alignment Loss}
For two hospitals $\mathcal{H}_i$ and $\mathcal{H}_j$ datasets, the Hospital-alignment loss averaged over all the $C$ classes, is calculated as,
\begin{align}
&\ell_\text{hospital alignment}(\mathcal{H}_i, \mathcal{H}_j; \psi', \theta') :=  \\ 
\nonumber
&~~~~~~~~~~\frac{1}{C} \sum_{c=1}^C \frac{1}{2} \Big[ \text{D}_\text{KL}(\bf{s}_c^{(i)} \| \bf{s}_c^{(j)}) + \text{D}_\text{KL}(\bf{s}_c^{(j)} \| \bf{s}_c^{(i)}) \Big], 
\end{align}
where $\text{D}_\text{KL}$ denotes the symmetrized Kullback–Leibler divergence, and $\bf{s}_c^{(\cdot)}$  denotes the soft confusion matrix which is calculated by applying softmax function with temperature $\tau > 1$ on the output of classification subnetwork, i.e. $S_{\theta}$.
\newline

\subsubsection{Triplet Loss}
Triplet loss \cite{schroff2015facenet, sikaroudi2021batch} is used for promoting hospital-independent class-specific cohesion and separation of instances. Each anchor, positive, and negative instances is denoted by $\bf{x}_a$, $\bf{x}_p$, and $\bf{x}_n$, respectively. 
For a batch of triplets, $\mathcal{\tau} := \{\bf{x}_a^b, \bf{x}_p^b, \bf{x}_n^b\}_{b=1}^B$, from all the source domain datasets $\{\mathcal{H}_k\}_{k=1}^K$, the average triplet loss is 
\begin{align}\label{equ:masf_triplet}
&\mathcal{L}_\text{triplet}(\mathcal{\tau}; \psi', \phi) := \\
\nonumber
&~~~\frac{1}{B} \sum_{b=1}^B \Big[\|M_{\phi}(G_{\psi'}(\bf{x}_a^b)) - M_{\phi}(G_{\psi'}(\bf{x}_p^b))\|_2^2 \nonumber   ~~- \\ 
\nonumber
&~~~~~~~~~~~~\|M_{\phi}(G_{\psi'}(\bf{x}_a^b)) - M_{\phi}(G_{\psi'}(\bf{x}_n^b))\|_2^2 + \alpha\Big]_+,
\end{align}
where $\|\cdot\|_2$ is the $\ell_2$ norm, $\alpha$ is a margin and $[\cdot]_+ := \max(\cdot,0)$.
\newline

\subsection{Gradient Updating}
First $\psi$ and $\theta$ weights of $S_{\theta}$ and $G_{\psi}$ are updated given by:
\begin{align}\label{equation_gradient_descent_cross_entropy}
(\psi', \theta') \gets (\psi, \theta) - \alpha \nabla_{\psi, \theta} \mathcal{L}_\text{ce}(\mathcal{H}_\text{tr}; \psi, \theta),
\end{align}
where $\alpha$ is the learning rate and $\nabla_{\psi, \theta}$ indicates the gradient with respect to $\psi$ and $\theta$ parameters. Then, the meta loss is calculated using weighted sum of hospital-alignment and triplet losses as 
\begin{align}
& \mathcal{L}_\text{meta}(\mathcal{H}_\text{tr}, \mathcal{H}_\text{te}, \mathcal{\tau}; \psi', \theta', \phi) \gets 
\nonumber \\ 
&~~~~~~~~~~~~~~~~~~~ \beta_1 \mathcal{L}_\text{hospital alignment}(\mathcal{H}_\text{tr}, \mathcal{H}_\text{te}; \psi', \theta') ~~~ + \nonumber \\ 
&~~~~~~~~~~~~~~~~~~~ \beta_2 \mathcal{L}_\text{triplet}(\mathcal{\tau}; \psi', \phi),
\label{meta_loss}
\end{align}
where $\beta_1$ and  $\beta_2$ are positive. 
After Eq. (\ref{equation_gradient_descent_cross_entropy}), two other gradient descent steps are done as 
\begin{eqnarray}
     (\psi, \theta) &\gets& (\psi, \theta) - \eta \nabla_{\psi, \theta} (\mathcal{L}_\text{ce} + \mathcal{L}_\text{meta}),
     \label{equation_gradient_descent_cross_entropy_and_meta}\\
\phi &\gets& \phi - \gamma \nabla_\phi \mathcal{L}_\text{triplet}(\mathcal{\tau}; \psi', \phi),
\label{equation_gradient_descent_triplet_loss_1}
\end{eqnarray}
where $\eta$ and $\gamma$ are the learning rates and $\phi$ is the $M_{\phi}$ subnetwork parameters. 
Eqs. (\ref{equation_gradient_descent_cross_entropy}), (\ref{equation_gradient_descent_cross_entropy_and_meta}), and (\ref{equation_gradient_descent_triplet_loss_1}) are repeated iteratively until convergence.

\RestyleAlgo{ruled}

\LinesNumbered

\begin{algorithm}[hbt!]
\caption{The Hospital-Agnostic Approach}\label{alg:hospital-agnostic}
\KwData{There are $K$ sets of WSIs: $\mathcal{H}_{s} = \{H_k\}^K_{k=1}$, and hyperparameters $\alpha,\eta,\gamma, \beta_1, \beta_2$}
\KwResult{There will be $K$ sets of feature extractor $G_\psi$, classifier $S_\theta$, metric embedding $M_\phi$ subnetworks}
\SetKwInOut{Input}{Preprocessing}
\Input{ \\
\footnotesize{
\begin{enumerate}
    \item Segmentation of tissues from the background using Otsu method
    \item Extracting $227 \times 227$ patches from the foreground regions
    \item Create $\mathcal{H}_{p} = \{H_k\}^K_{k=1}$ using the patches
\end{enumerate}
}
}
\ForEach{ $k$, \text{splits} $\mathcal{H}_p$ \text{into} \(\mathcal{H}_{\text{external}} =  \{H_k\}\) \text{and} \(\mathcal{H}_{\text{internal}} =  \{H_i\}^K_{i\neq k}\)}{
    \Repeat{convergence}{
    Randomly split \(\mathcal{H}_{\text{internal}}\) into disjoint meta-train \(\mathcal{H}_{tr}\)  and meta-test \(\mathcal{H}_{te}\), \\
    Update using Equation \ref{equation_gradient_descent_cross_entropy},    \\
    \text{Compute hospital alignment loss:} \\
     \pushline \nonl 
     \(\mathcal{L}_\textrm{\scriptsize{hospital alignment}} \gets \frac{1}{|\mathcal{H}_\text{tr}|}  \sum_{\frac{1}{|\mathcal{H}_\text{te}|}\mathcal{H}_i \in \mathcal{H}_\text{tr}} \sum_{\mathcal{H}_j \in \mathcal{H}_\text{te}}\) \(\ell_\textrm{\scriptsize{hospital alignment}} (\mathcal{H}_i, \mathcal{H}_j; \psi', \theta')\),\\ 
    \popline Compute triplet loss using Equation \ref{equ:masf_triplet}, \\
    Compute meta loss using Equation \ref{meta_loss}, \\
    Update using Equation \ref{equation_gradient_descent_cross_entropy_and_meta},\\
    Update using Equation \ref{equation_gradient_descent_triplet_loss_1},
    }
}
\end{algorithm}

\section{EXPERIMENTS}

\subsection{Dataset}

\textbf{Renal Cell Carcinoma} ---
The most prevalent kidney cancer in adults is Renal Cell Carcinoma (RCC)  which is a heterogeneous group of diseases with different morphology, molecular characteristics, clinical outcomes, and treatment responses. Since RCCs are classified based on their histological subtypes, classification is a critical diagnostic task and it is conducive to a higher chance of successful treatment. Hence, RCC classification is our case study.

\textbf{WSI repository} ---
All WSIs and clinical information were downloaded from The Cancer Genome Atlas (TCGA) data portal. From the whole dataset, some of the WSIs were removed due to reading and compatibility issues. Furthermore, only diagnostic slides scanned at 40x magnification have been considered. In this study, the trial sites with sufficiently large number ($\gtrapprox 30$) of WSIs across all three subtypes (ccRCC, pRCC, crRCC) were considered. In TCGA, only three sites, namely NCI Urologic Center, International Genomics Center, and Memorial Sloan Kettering Cancer Center (MSKCC) met this criterion. A fourth center has been formed by combining the Harvard and MD Anderson repositories. Overall, we used 467 RCC WSIs of TCGA.

\textbf{Details} --- Each WSI repository has been split into $45\%$, $45\%$, $10\%$ chunks (similar to \cite{Li2017dg}), respectively, for training, validation, and testing. Then, the foreground regions for each WSI, i.e., the tissue, are extracted. Following that, further morphological closing is performed to fill up tiny gaps and holes of each WSI. Then, for the sake of compatibility with backbone input size (AlexNet), $227\times227\times3$ RGB patches were extracted from the segmented tissue of each WSI at 40x magnification with no overlap. The patches containing more than 50\% of the background region were removed. In addition, re-sampling of all the extracted patches was applied to form a balanced dataset. For each of the trial sites and subtypes in $\mathcal{H}_p$, approximately 70,000 patches were sampled.

\begin{figure*}
\begin{tabular}{ccc}
    \includegraphics[width=0.3\linewidth]{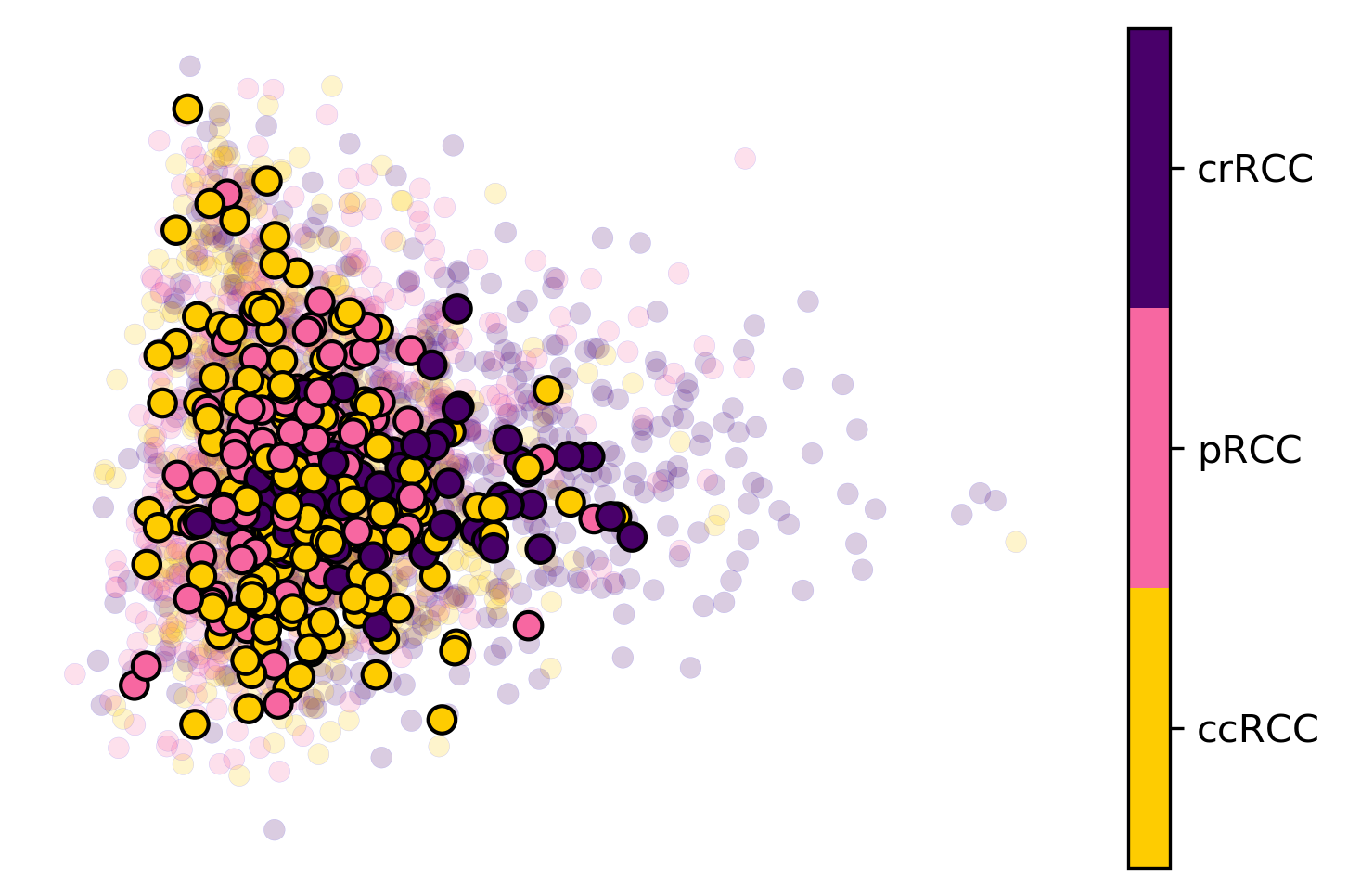} &
    \includegraphics[width=0.3\linewidth]{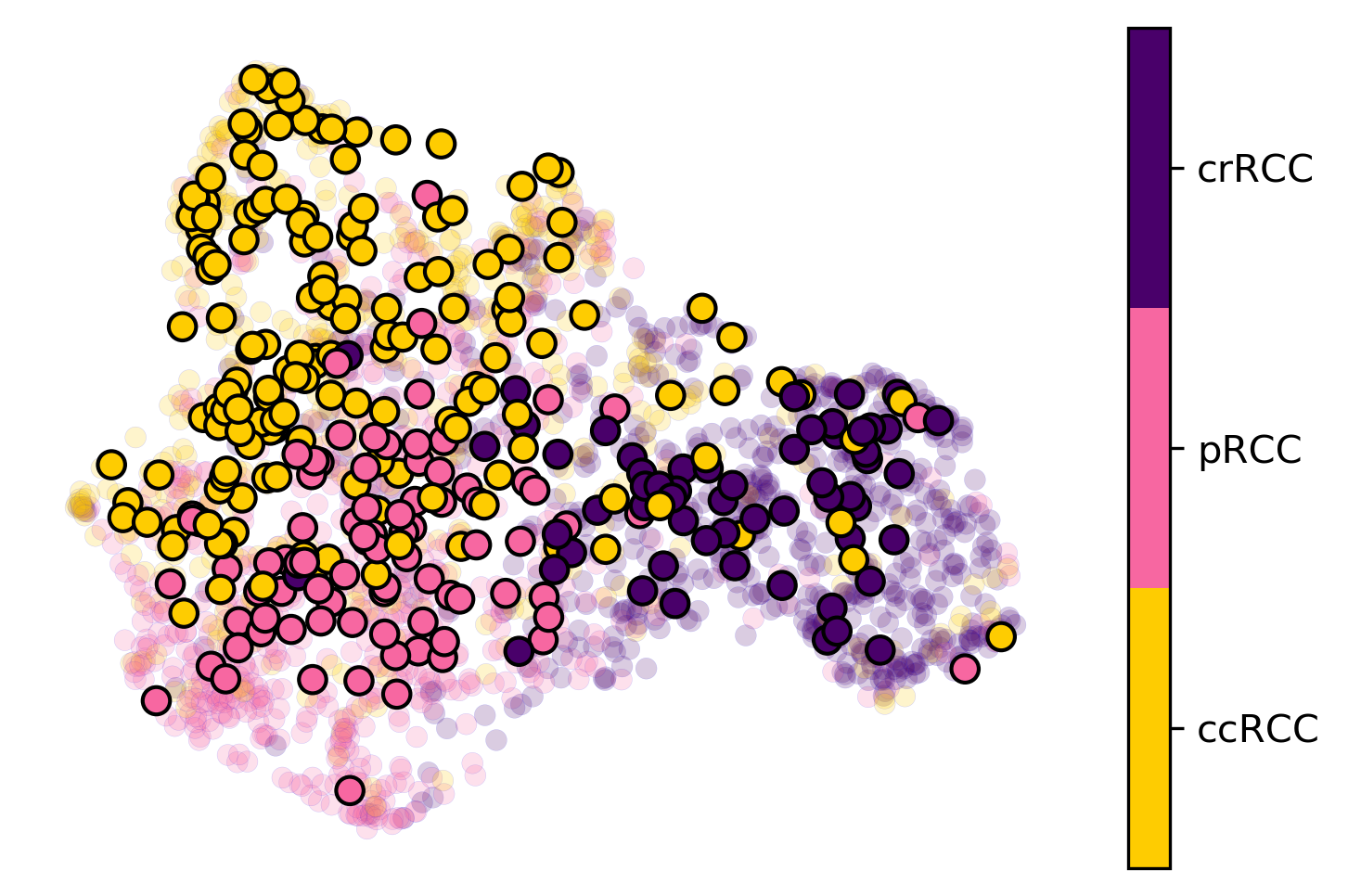} &
    \includegraphics[width=0.3\linewidth]{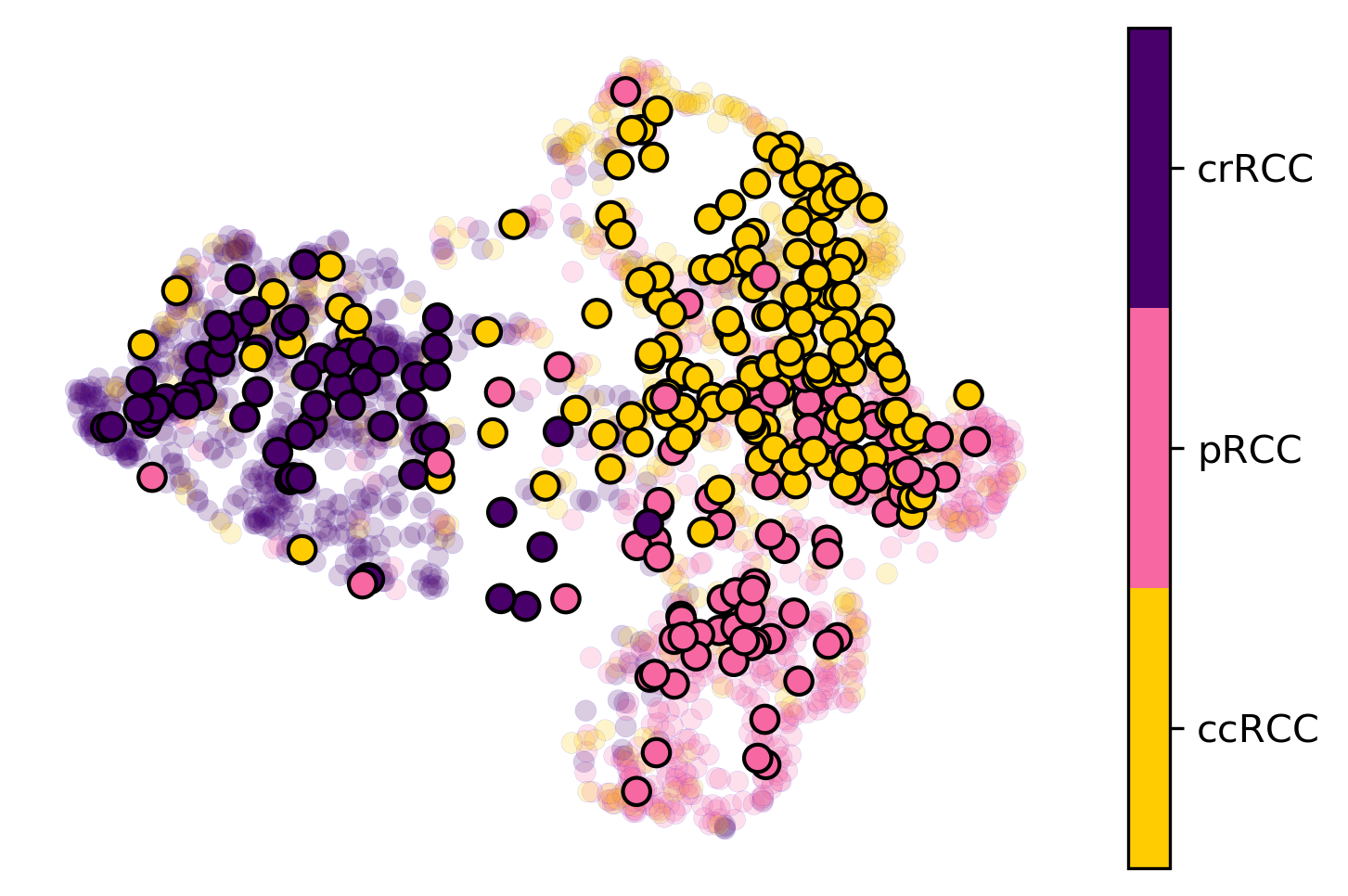} \\
(a) & (b) & (c) \\[6pt]
\end{tabular}
\caption{(a) ImageNet pre-trained AlexNet, (b) Baseline, (c) Fine-tuned using hospital-agnostic learning regime. The hold-out trial site is ``\textbf{NCI Urologic Oncology Branch}''. Note that the resulting 2-dimensional representations have been transparently visualized for each patch representation by its ground truth slide-level label. The 2-dimensional representations of all patches were aggregated (averaged) for each WSI to attain the slide-level representations which are shaded opaque with a dark border.}
\label{fig:visualization_nci}
\end{figure*}

\subsection{Experimental Setup}
\textbf{Baseline} --
The baseline method in the experiments is to fine-tune the backbone using conventional cross-entropy loss. To have this, all WSIs of the trial sites were merged and $G_{\psi}\circ S_{\theta}$ was trained by standard supervised-learning on $\mathcal{L}_{\text{ce}}$ with the same hyperparameters as the hospital-agnostic regime.

\textbf{Backbone}---
Since the proposed approach is a model-agnostic (consequently architecture-agnostic) approach, the general architecture stays a generic concept. In other words, the idea can be applied to other architectures. In this study, ImageNet-pretrained ``\emph{AlexNet}'' is adopted as the backbone.

\textbf{Hyperparameter and Details
}---
The Adam optimizer \cite{kingma2014adam}, initialized with the learning rate $10^{-3}$, was employed for optimization. By stacking two fully-connected layers with output sizes of 1,024 and 256, the metric loss subnetwork $M_{\phi}$ is connected to the last fully-connected layer. The triplet loss was utilized with $\beta_{2} = 0.005$, $\beta_{1} = 1.0$ for the $\mathcal{L}_{\text{triplet}}$ such that it is in a similar scale to $\mathcal{L}_{\text{ce}}$ and $\mathcal{L}_{\text{hospital alignment}}$. For the inner optimization, the gradients that had a norm greater than a threshold\footnote{Similar to \cite{dou2019domain}, the threshold was set to $2.0$} were clipped to prevent them from exploding, as this step employs simple non-adaptive gradient descent (with learning rate $\alpha = 1e-5$). As well, an Adam optimizer is used to do meta-updates with a $\eta =10^{-5}$ as learning rate without decay, and the batch size was $432$ patches. The hyperparameter for the metric-learning margin was set heuristically (according to \cite{dou2019domain}) to be $10$ by examining the distances between clusters of class characteristics, and the metric loss learning rate was set to $\gamma = 10^{-5}$ with a  maximum number of iterations of $1,000$.

\subsection{Results}
In this study, for each trial site, i.e. $\mathcal{H}_{\text{external}}$, using leave-one-\emph{hospital}-out, the backbone is only trained using the rest of the repositories, i.e. $\mathcal{H}_{\text{internal}}$.

\textbf{Low-dimensional Embedding Visualization
}--- Once the training is done, the model is used to get a 4,096-dimensional latent space representation for every patch in the corresponding hold-out test set.  According to the best practice in the literature \cite{sakaue2020dimensionality}, the first 20 principal components were achieved using PCA \cite{wold1987principal} and then UMAP \cite{mcinnes2018umap} was applied to lower the dimension to 2. As per Fig. \ref{fig:visualization_nci} pre-trained AlexNet weights have not been able to differentiate RCC subtypes. While the hospital-agnostic learning regime has outperformed the baseline in terms of having a discriminative embedding space. These results were replicated for other hold-out trial sites in our experiments.

\textbf{RCC Classification Accuracy
}---
Once the training is done, the slide-level accuracy is calculated. For this, the softmax outputs, which are the probability scores representing the degree of belongingness to each of the RCC subtypes (ccRCC, pRCC, and crRCC), were averaged over the patches of each WSI. Hence, the predicted subtype for each WSI could be assigned. According to Table \ref{table:acc}, the proposed hospital-agnostic learning regime has outperformed the baseline when the hold-out trial site were ``NCI Urologic Oncology Branch'', ``Harvard ad MD Andersson'', and  ``MSKCC''. A case in point is when the ``Harvard and MD Anderson'' was the trial site. As far as, this repository has been formed by combining two other different trial sites, the domain shift is expectedly more significant. As per Table. \ref{table:acc}, the hospital-agnostic learning regime has been able to outperform the baseline, i.e., $\approx7\%$ when the hold-out repository was ``Harvard and MD Anderson".

\begin{table}[]
\caption{Slide-level accuracy for different trial sites.}

\begin{tabular}{lcc}

\hline
                                      & \multicolumn{2}{c}{Accuracy(\%)}                              \\ \cline{2-3} 
{\textbf{Hold-out Trial Site}} & Hospital-Agnostic             & Baseline                      \\ \hline
International Genomics Consortium     & 79.31                         & \cellcolor{green!25}\textbf{80.45} \\ \hline
Harvard and MD Anderson               & \cellcolor{green!25}\textbf{79.31} & 72.41                         \\ \hline
MSKCC                                 & \cellcolor{green!25}\textbf{82.65} & 81.18                         \\ \hline
NCI Urologic Oncology Branch          & \cellcolor{green!25}\textbf{84.09}  & 81.81                         \\ \hline
\end{tabular}
\label{table:acc}
\end{table}

\section{Conclusion}
We proposed an effective hospital-agnostic learning regime. The approach has been inspired by a DG technique, namely MASF, in which a DNN is trained by three loss functions in an episodic learning regime. These loss functions are cross-entropy used for hard class separation, triplet loss for soft class separation, and the KL divergence for hospital alignment. By leveraging this learning regime, a hospital-agnostic model was trained, and its effectiveness in achieving a discriminative latent space representation and classification of RCC subtypes were demonstrated and analyzed using low-dimensional embedding visualization and classification accuracy in comparison to conventional supervised learning.

\vspace{0.05in}
\textbf{Acknowledgment} Images are from TCGA repository.   

\bibliographystyle{IEEEtran}
\bibliography{main}

\begin{thebibliography}{10}
\providecommand{\url}[1]{#1}
\csname url@samestyle\endcsname
\providecommand{\newblock}{\relax}
\providecommand{\bibinfo}[2]{#2}
\providecommand{\BIBentrySTDinterwordspacing}{\spaceskip=0pt\relax}
\providecommand{\BIBentryALTinterwordstretchfactor}{4}
\providecommand{\BIBentryALTinterwordspacing}{\spaceskip=\fontdimen2\font plus
\BIBentryALTinterwordstretchfactor\fontdimen3\font minus
  \fontdimen4\font\relax}
\providecommand{\BIBforeignlanguage}[2]{{%
\expandafter\ifx\csname l@#1\endcsname\relax
\typeout{** WARNING: IEEEtran.bst: No hyphenation pattern has been}%
\typeout{** loaded for the language `#1'. Using the pattern for}%
\typeout{** the default language instead.}%
\else
\language=\csname l@#1\endcsname
\fi
#2}}
\providecommand{\BIBdecl}{\relax}
\BIBdecl

\bibitem{chen2014histological}
Z.~Chen, D.~Shin, S.~Chen, K.~Mikhail, O.~Hadass, B.~N. Tomlison, D.~Korkin,
  C.-R. Shyu, J.~Cui, D.~C. Anthony \emph{et~al.}, ``Histological quantitation
  of brain injury using whole slide imaging: a pilot validation study in
  mice,'' \emph{PLoS One}, vol.~9, no.~3, p. e92133, 2014.

\bibitem{jimenez2017analysis}
O.~Jimenez-del Toro, S.~Ot{\'a}lora, M.~Andersson, K.~Eur{\'e}n, M.~Hedlund,
  M.~Rousson, H.~M{\"u}ller, and M.~Atzori, ``Analysis of histopathology
  images: From traditional machine learning to deep learning,'' in
  \emph{Biomedical Texture Analysis}.\hskip 1em plus 0.5em minus 0.4em\relax
  Elsevier, 2017, pp. 281--314.

\bibitem{tizhoosh2018artificial}
H.~R. Tizhoosh and L.~Pantanowitz, ``Artificial intelligence and digital
  pathology: challenges and opportunities,'' \emph{Journal of pathology
  informatics}, vol.~9, 2018.

\bibitem{ruder2017transferlearning}
S.~Ruder, ``{Transfer Learning - Machine Learning's Next Frontier},''
  \url{http://ruder.io/transfer-learning/}, 2017.

\bibitem{sikaroudi2020supervision}
M.~Sikaroudi, A.~Safarpoor, B.~Ghojogh, S.~Shafiei, M.~Crowley, and H.~R.
  Tizhoosh, ``Supervision and source domain impact on representation learning:
  A histopathology case study,'' in \emph{Int. Conf. of the IEEE Eng. in
  Medicine \& Biology Society (EMBC)}.\hskip 1em plus 0.5em minus 0.4em\relax
  IEEE, 2020, pp. 1400--1403.

\bibitem{torralba2011unbiased}
A.~Torralba and A.~A. Efros, ``Unbiased look at dataset bias,'' in \emph{CVPR
  2011}.\hskip 1em plus 0.5em minus 0.4em\relax IEEE, 2011, pp. 1521--1528.

\bibitem{quionero2009dataset}
J.~Quionero-Candela, M.~Sugiyama, A.~Schwaighofer, and N.~D. Lawrence,
  \emph{Dataset shift in machine learning}.\hskip 1em plus 0.5em minus
  0.4em\relax The MIT Press, 2009.

\bibitem{yagi2011color}
Y.~Yagi, ``Color standardization and optimization in whole slide imaging,'' in
  \emph{Diagnostic pathology}, vol.~6.\hskip 1em plus 0.5em minus 0.4em\relax
  Springer, 2011, p. S15.

\bibitem{dehkharghanian2021biased}
T.~Dehkharghanian, A.~A. Bidgoli, A.~Riasatian, P.~Mazaheri, C.~J. Campbell,
  L.~Pantanowitz, H.~Tizhoosh, and S.~Rahnamayan, ``Biased data, biased ai:
  Deep networks predict the acquisition site of tcga images,'' 2021.

\bibitem{reinhard2001color}
E.~Reinhard, M.~Adhikhmin, B.~Gooch, and P.~Shirley, ``Color transfer between
  images,'' \emph{IEEE Computer graphics and applications}, vol.~21, no.~5, pp.
  34--41, 2001.

\bibitem{dou2019domain}
Q.~Dou, D.~C. de~Castro, K.~Kamnitsas, and B.~Glocker, ``Domain generalization
  via model-agnostic learning of semantic features,'' in \emph{Advances in
  Neural Info. Proc. Syst.}, 2019, pp. 6450--6461.

\bibitem{sikaroudi2021magnification}
M.~Sikaroudi, B.~Ghojogh, F.~Karray, M.~Crowley, and H.~R. Tizhoosh,
  ``Magnification generalization for histopathology image embedding,'' in
  \emph{2021 IEEE 18th International Symposium on Biomedical Imaging
  (ISBI)}.\hskip 1em plus 0.5em minus 0.4em\relax IEEE, 2021, pp. 1864--1868.

\bibitem{bilen2017universal}
H.~Bilen and A.~Vedaldi, ``Universal representations: The missing link between
  faces, text, planktons, and cat breeds,'' \emph{arXiv preprint
  arXiv:1701.07275}, 2017.

\bibitem{finn2017model}
C.~Finn, P.~Abbeel, and S.~Levine, ``Model-agnostic meta-learning for fast
  adaptation of deep networks,'' in \emph{The 34th International Conference on
  Machine Learning}, 2017.

\bibitem{otsu1979threshold}
N.~Otsu, ``A threshold selection method from gray-level histograms,''
  \emph{IEEE transactions on systems, man, and cybernetics}, vol.~9, no.~1, pp.
  62--66, 1979.

\bibitem{schroff2015facenet}
F.~Schroff, D.~Kalenichenko, and J.~Philbin, ``Facenet: A unified embedding for
  face recognition and clustering,'' in \emph{Proceedings of the IEEE
  Conference on Computer Vision and Pattern Recognition}, 2015, pp. 815--823.

\bibitem{sikaroudi2021batch}
M.~Sikaroudi, B.~Ghojogh, F.~Karray, M.~Crowley, and H.~R. Tizhoosh,
  ``Batch-incremental triplet sampling for training triplet networks using
  bayesian updating theorem,'' in \emph{2020 25th International Conference on
  Pattern Recognition (ICPR)}.\hskip 1em plus 0.5em minus 0.4em\relax IEEE,
  2021, pp. 7080--7086.

\bibitem{Li2017dg}
D.~Li, Y.~Yang, Y.-Z. Song, and T.~Hospedales, ``Deeper, broader and artier
  domain generalization,'' in \emph{International Conference on Computer
  Vision}, 2017.

\bibitem{kingma2014adam}
D.~P. Kingma and J.~Ba, ``Adam: A method for stochastic optimization,''
  \emph{arXiv preprint arXiv:1412.6980}, 2014.

\bibitem{sakaue2020dimensionality}
S.~Sakaue, J.~Hirata, M.~Kanai, K.~Suzuki, M.~Akiyama, C.~L. Too, T.~Arayssi,
  M.~Hammoudeh, S.~Al~Emadi, B.~K. Masri \emph{et~al.}, ``Dimensionality
  reduction reveals fine-scale structure in the japanese population with
  consequences for polygenic risk prediction,'' \emph{Nature communications},
  vol.~11, no.~1, pp. 1--11, 2020.

\bibitem{wold1987principal}
S.~Wold, K.~Esbensen, and P.~Geladi, ``Principal component analysis,''
  \emph{Chemometrics and intelligent laboratory systems}, vol.~2, no. 1-3, pp.
  37--52, 1987.

\bibitem{mcinnes2018umap}
L.~McInnes, J.~Healy, and J.~Melville, ``Umap: Uniform manifold approximation
  and projection for dimension reduction,'' \emph{arXiv preprint
  arXiv:1802.03426}, 2018.

\end{thebibliography}

\end{document}